\documentclass[aps]{revtex4}
\usepackage{graphicx}
\usepackage{amsmath}
\usepackage{amsfonts}

\begin{document}

\title{A Geometrical Interpretation of Grassmannian Coordinates}
\author{V. Dzhunushaliev}
\email{dzhun@hotmail.kg}
\affiliation{Dept. Phys.
and Microel. Engineer., Kyrgyz-Russian
Slavic University, Bishkek, Kievskaya Str. 44, 720000, Kyrgyz
Republic}

\begin{abstract}
A geometrical interpretation of Grassmannian anticommuting
coordinates is given. They are taken to represent an
indefiniteness inherent in every spacetime point on the level of
the spacetime foam. This indeterminacy is connected with the fact
that in quantum gravity in some approximation we do not know the
following information : are two points connected by a quantum
wormhole or not ? It is shown that: (a) such indefiniteness can be 
represented by Grassmanian numbers, (b) a displacement of the 
wormhole mouth is connected with a change  of the Grassmanian 
numbers (coordinates). In such an interpretation of supersymmetry the
corresponding supersymmetrical fields must be described in an
invariant manner on the background of the spacetime foam.
\end{abstract}
\maketitle

\section{Introduction}

The idea of superspace enlarges the spacetime points labelled by
the coordinates $x^\mu$ by adding two plus two anticommuting
Grassmannian coordinates $\theta_\alpha$ and
$\bar\theta_{\dot{\beta}}$ ($\alpha$ and $\dot\beta$ are the spinor 
indices). 
Thus the coordinates on superspace are
$Z^M = (x^\mu, \theta_\alpha, \bar\theta_{\dot{\beta}})$ and for
brevity we introduce $\theta = (\theta_\alpha,
\bar{\theta}_{\dot{\beta}})$. Ordinary anticommuting coordinates
$\theta$ are abstract Grassmannian numbers. Nevertheless the
following question should be asked : can the Grassmannian
coordinates have some physical meaning~? In this paper I would
like to show that a physical meaning can be given in terms of an
indefiniteness inherent at every point of spacetime.
\par
\begin{figure}
\begin{center}
\fbox{
\includegraphics[height=5cm,width=5cm]{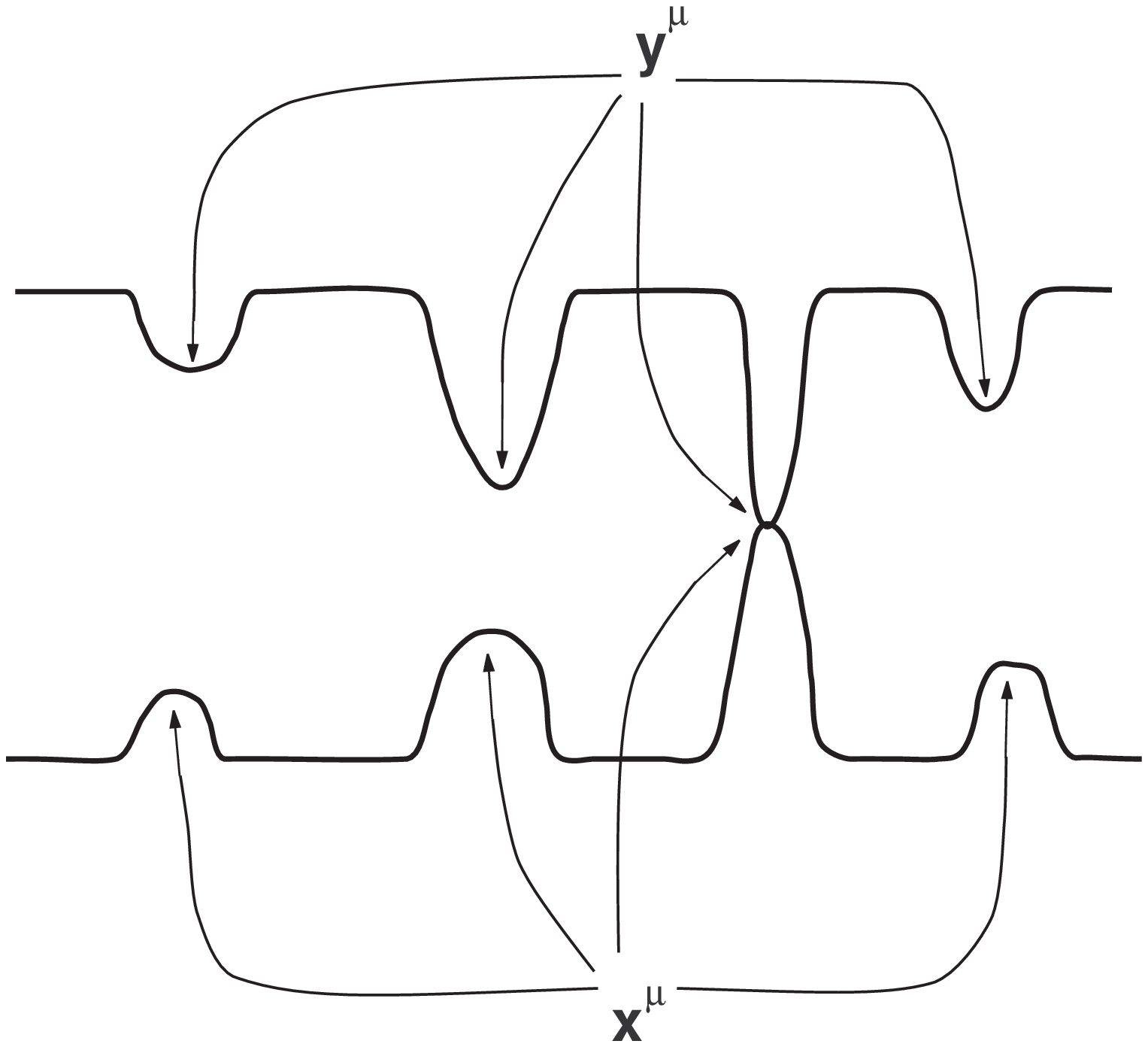}}
\caption{The wormhole in the spacetime foam
as an identification of two points $y^\mu$ and
$x^\mu$. We shall name such wormhole as in Ref. \cite{smolin} : 
a minimalist wormhole. Such minimalist wormhole is an approximation of the 
real wormhole in which we neglect all linear sizes (longitudinal and 
cross section).}
\label{fig1}
\end{center}
\end{figure}

What kind of indefiniteness can be incorporated in spacetime ?
We assume that such an indeterminacy can appear in quantum
gravity on the level of the spacetime foam. The notion of the
spacetime foam was introduced by Wheeler for describing the
possible complex structure of spacetime on the Planck scale
\cite{wheeler}.
It is postulated that the spacetime foam is a cloud of quantum
wormholes with a typical linear size of the Planck length. Schematically
in some rough approximation we can imagine the appearing/disappearing
of quantum wormholes as pasting together two points with a
subsequent break (see, Fig.\ref{fig1}). In our approach we
deliberately neglect the following information~:
\textbf{\textit{whether two points $y^\mu$ and $x^\mu$
are connected by a quantum wormhole or not.}}
In such an approximation we have an indeterminacy for each spacetime
point. We can say that in this approximation the spacetime foam in quantum
gravity is described in some effective manner~: quantum minimalist wormholes are
approximated as an indefiniteness inherent in every spacetime point.

\section{Physical idea}

\begin{figure}
\begin{center}
\fbox{
\includegraphics[height=5cm,width=8cm]{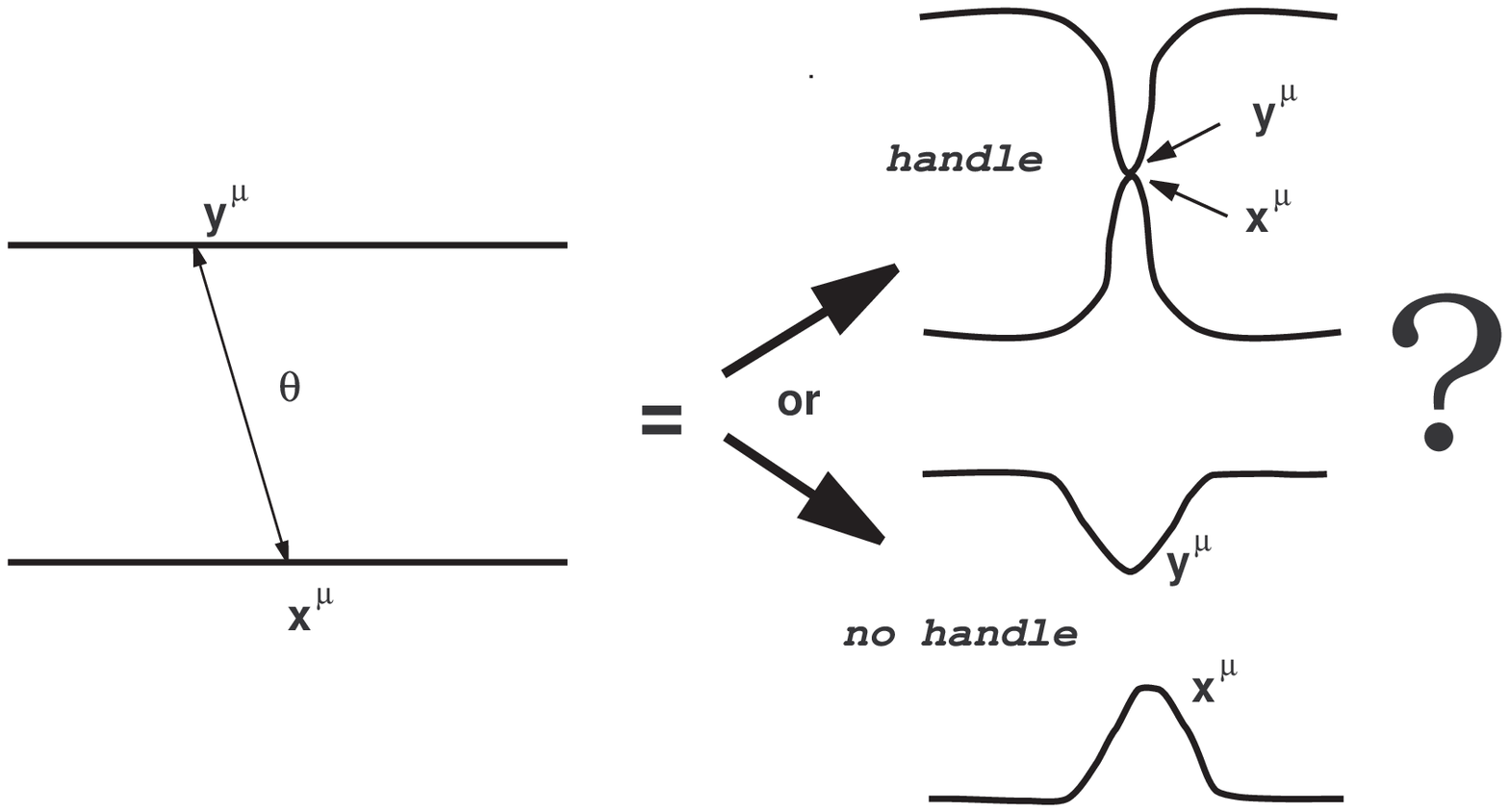}}
\caption{A geometrical interpretation of the Grassmannian
coordinate $\theta$. The Grassmannian number $\theta$ describes
the indefiniteness (the lack of knowledge) : are the points
$y^\mu$ and $x^\mu$ identified or separated ?} \label{fig3}
\end{center}
\end{figure}

Our assumption is that Grassmannian coordinate $\theta$ describes
the indefiniteness (the loss of information) 
of our knowledge about two points $x^\mu$ and
$y^\mu$~: we do not know if these points are connected by a
quantum minimalist wormhole or not (see, Fig.\ref{fig3})
\begin{equation}
  \theta = \left(\theta_\alpha , \bar\theta_{\dot{\beta}}\right) =
  Id\left( x^{\alpha\dot{\beta}} \;
  \raisebox{1.5ex}{$\underleftrightarrow{
  \scriptstyle{\;\;yes \; or \; no \; ?\;\;}}$}
  \; y^{\alpha\dot{\beta}}
  \right)
\label{sec1-10}
\end{equation}
where $Id (\ldots)$ is the designation for the identification
procedure; $x_{\alpha\dot{\beta}} =
\sigma_{\mu\alpha\dot{\beta}} x^\mu$;
$\sigma^\mu_{\alpha\dot{\beta}} = \{\textbf{1}, -\sigma^i\}$; $\sigma^i$ are
the Pauli matrices; $\alpha,\dot{\beta} = 1,2$ are the matrix indices;
$\mu = 0,1,2,3$ is the spacetime index; $i = 1,2,3$ is the space index
\begin{eqnarray}
  x^{\alpha\dot{\beta}} & = & \varepsilon^{\alpha\gamma}
  \varepsilon^{\dot{\beta}\dot{\delta}} x_{\gamma\dot{\delta}} ,
  \label{sec1-11}\\
  \varepsilon^{\alpha\beta} & = & \varepsilon^{\dot{\alpha}\dot{\beta}} =
    \begin{pmatrix}
    0 & 1 \\
    -1 & 0
  \end{pmatrix} ,
  \label{sec1-11a} \\
  \varepsilon_{\alpha\beta} & = & \varepsilon_{\dot{\alpha}\dot{\beta}} =
    \begin{pmatrix}
    0 & -1 \\
    1 & 0
  \end{pmatrix} .
\label{sec1-13}
\end{eqnarray}
\par
Such an interpretation of $\theta$ is similar to
the interpretation of spin : in certain situations we do not know
the value of the $z-$~projection of spin $(\hbar/2$ or $-\hbar/2)$.
In Pauli's words spin is ``a non-classical two-valuedness'' or in the
context of this paper it is an ``indefiniteness'' (``indeterminacy''). In
modern terms ``non-classical two-valuedness'' is spin and
it can be described only using spinors.
\par
This comparison allows us to postulate that the
above-mentioned indeterminacy connected
with the spacetime foam
(pasting together/cutting off of points) can be described using spinors.
Let us introduce the scalar product for spinors $\psi$ and $\chi$ as
(see, for example, Ref. \cite{bilal})
\begin{equation}
  \psi\chi = \psi^\alpha \chi_\alpha =
  \varepsilon^{\alpha\beta} \psi_\beta \chi_\alpha .
\label{sec1-14}
\end{equation}
For the scalar product should imply
\begin{equation}
  \psi\chi = \chi\psi
\label{sec1-15}
\end{equation}
but
\begin{equation}
  \chi\psi = \varepsilon^{\alpha\beta} \chi_\beta \psi_\alpha
\label{sec1-16}
\end{equation}
hence
\begin{equation}
  \varepsilon^{\alpha\beta} \psi_\beta \chi_\alpha =
  -\varepsilon^{\alpha\beta} \chi_\alpha \psi_\beta
\label{sec1-17}
\end{equation}
this means that
\begin{equation}
  \psi_\beta \chi_\alpha = -\chi_\alpha \psi_\beta ,
\label{sec1-18}
\end{equation}
\textit{i.e.} the components of a spinor describing the
above-mentioned indefiniteness are anticommuting Grassmannian
numbers. These arguments allow us to say that \textbf{\textit{the
indefiniteness can be described using anticommuting Grassmannian
coordinates.}}

\section{Spacetime foam and indefiniteness}

\subsection{Precedent results}

In this subsection I would like to remind some definitions 
from Ref. \cite{smolin}. Let us define 
an operator $\hat{A}(x,y)$ : it is an operator which identifies 
two points $x$ and $y$ or another words this operator creates 
a minimalist wormhole (see Fig. \ref{fig1}). It makes no sense to 
identify these points $(x,y)$ twice, therefore we have the following 
property 
\begin{equation}
  \left(\hat{A}(x,y)\right)^2 = 0.
\label{sec3.1-1}
\end{equation}
Let us introduce Weyl fermion $\psi^\alpha(x)$ with the standard 
anticommuting relations 
\begin{equation}
  \left[
  \psi^\alpha (x) , \psi^\dag _\beta (y) 
  \right] = \delta^\alpha_\beta \delta^3(x - y) .
\label{sec3.1-2}
\end{equation}
In the context of Ref. \cite{smolin} 
we have the following relation between minimalist wormhole and Weyl 
fermion 
\begin{equation}
  \psi^\alpha (x) \psi^\beta (y) \Longleftrightarrow 
  -\varepsilon^{\alpha\beta} \hat{A}(x,y) .
\label{sec3.1-3}
\end{equation}
Let us introduce an operator $\hat{B}(x \rightarrow x')$ 
which moves a mouth of a wormhole from point $x$ to point $x'$ 
\begin{equation}
  \hat{B}(x \rightarrow x') \hat{A}(y, x) = 
  \hat{A}(y, x') .
\label{sec3.1-4}
\end{equation}

\subsection{Mathematical definitions}

In this subsection I would like to connect a modification of 
the operators $\hat{A}(x,y)$ and $B(x \rightarrow x')$ 
introduced by Smolin with the Grassmanian coordinates. 
\par 
Let we have 
an operator $\hat{A}^{AB}(x,y)$ describing a quantum 
state in which the space with two points $x$ and $y$ fluctuates 
between two possibilities : points $(x,y)$ \textbf{either} identified 
\textbf{or} not.   
Let the operator $\hat{A}^{AB}(x,y)$ 
like to the previous Smolin's definition) 
has the following property 
\begin{equation}
    \hat{A}^{AB}(x,y) \hat{A}_{AB}(x,y) = 0 .
    \label{sec3.2-1}
\end{equation}
We should note that 
\begin{itemize}
  \item 
  we do not here define the indices $(A,B)$;
  \item 
  $\hat{A}_{AB}(x,y) = \epsilon_{AC}\epsilon_{BD}
  \hat{A}^{CD}(x,y)$ and the matrixes $\epsilon_{AB}$ we will define 
  later.  
\end{itemize}
The solution of Eq.\eqref{sec3.2-1} we search in the form 
\begin{equation}
    \hat{A}^{AB}(x,y) = \theta^A(x) \theta^B(y).
\label{sec3.2-2}
\end{equation}
From Eq's \eqref{sec3.2-1} and \eqref{sec3.2-2} we have 
\begin{equation}
  \theta^A(x) \theta^B(y) \theta_A(x) \theta_B(y) = 0.
\label{sec3.2-3}
\end{equation}
This equation has the following two simplest solution. 
The first is 
\begin{eqnarray}
    A & = & \alpha, \; B = \beta , \; (\alpha , \beta= 1,2) ,
    \label{sec3.2-0a}\\
    \epsilon_{\alpha\beta} & = & \varepsilon_{\alpha\beta}, 
    \label{sec3.2-0b}\\
    \hat{A}^{\alpha\beta}(x,y) & = & \varepsilon^{\alpha\beta} 
    \hat{A}(x,y) = \theta^\alpha \theta^\beta .
    \label{sec3.2-0c}
\end{eqnarray}
It means that $\theta^\alpha$ is an undotted spinor of 
$(\frac{1}{2},0)$ representation of Sl(2,\textbf{C}) group. 
\begin{eqnarray}
    \theta^\alpha(x) & = & \theta^\alpha(y) = \theta^\alpha = const ,
    \label{sec3.2-4}\\
    \theta^\alpha \theta^\beta & = & - \theta^\beta \theta^\alpha ; 
    \qquad \alpha \neq \beta ,
    \label{sec3.2-5}\\
    \left(\theta^\alpha \right)^2 & = & 0 .
    \label{sec3.2-6}
   \end{eqnarray}
The second solution is the same but only with the replacement 
$\alpha \rightarrow \dot \alpha$ and 
$\beta \rightarrow \dot \beta$ 
\begin{eqnarray}
    A & = & \dot\alpha, \; B = \dot\beta , \; 
    (\dot\alpha , \dot\beta= 1,2) , 
    \label{ces3.2-7}\\
    \epsilon_{\dot\alpha\dot\beta} & = & 
    \varepsilon_{\dot\alpha\dot\beta} ,
    \label{sec3.2-7}\\
    \hat{A}^{\dot\alpha\dot\beta}(x,y) & = & 
    \varepsilon^{\dot\alpha\dot\beta} 
    \hat{A}(x,y) = \theta^{\dot\alpha} \theta^{\dot\beta} ,
    \label{sec3.2-7a}\\
    \bar\theta^{\dot\alpha}(x) & = & \bar\theta^{\dot\alpha}(y) = 
    \bar\theta^{\dot\alpha} = const ,
    \label{sec3.2-8}\\
    \bar\theta^{\dot\alpha} \bar\theta^{\dot\beta} & = & 
    - \bar\theta^{\dot\beta} \bar\theta^{\dot\alpha} ; 
    \qquad \dot\alpha \neq \dot\beta ,
    \label{sec3.2-9}\\
    \left(\bar\theta^{\dot\alpha} \right)^2 & = & 0 .
    \label{sec3.2-10}
\end{eqnarray}
In this case $\bar\theta^{\dot\alpha}$ is a dotted spinor of 
$(0,\frac{1}{2})$ representation. More concretely we have 
\begin{eqnarray}
    \theta^1 \theta^2 & = & - \theta^2 \theta^1 
    \quad \text{or} \quad
    \bar\theta^{\dot 1} \bar\theta^{\dot 2} = 
    - \bar\theta^{\dot 2} \bar\theta^{\dot 1} , 
    \label{sec3.2-11}\\
    \left( \theta^1 \right)^2 & = & 
    \left( \theta^2 \right)^2 = 
    \left( \bar\theta^{\dot 1} \right)^2 = 
    \left( \bar\theta^{\dot 2} \right)^2 = 0 .
    \label{sec3.2-12}
\end{eqnarray}
Such two-valuedness forces us to introduce both possibilities : 
$\theta = \{ \theta^\alpha , \bar\theta^{\dot\alpha} \}$. 
\par 
Now we would like to introduce an infinitesimal operator 
$\delta\hat{B}(x^\mu \rightarrow {x'}^\mu)$ 
(like to Smolin) of a displacement of the 
wormhole mouth 
\begin{eqnarray}
    \delta\hat{B}(x^\mu & \rightarrow & {x'}^\mu) 
    \hat{A}^{\gamma\delta}(y^\mu , x^\mu) = 
    \hat{A}^{\gamma\delta}(y^\mu , {x'}^\mu ) ,
    \label{sec3.2-12a}\\
    \hat{A}^{\gamma\delta}(y^\mu , x^\mu) & = & 
    \theta^\gamma \theta^\delta ,
    \label{sec3.2-12b}\\
    \hat{A}^{\gamma\delta}(y^\mu , {x'}^\mu) & = & 
    {\theta '}^\gamma {\theta '}^\delta  = 
    \left( \theta^\gamma + \varepsilon^\gamma \right)
    \left( \theta^\delta + \varepsilon^\delta \right) 
    \approx  
    \nonumber \\
    \theta^\gamma \theta^\delta & + & 
    \varepsilon^\gamma \theta^\delta - 
    \varepsilon^\delta \theta^\gamma 
    \label{sec3.2-12c}
\end{eqnarray}
here $\varepsilon^\alpha$ is an infinitesimal Grassmannian number. 
Therefore we have the following equation for the definition of 
$\delta\hat{B}(x^\mu \rightarrow {x'}^\mu)$ operator 
\begin{equation}
    \delta\hat{B}\left(x^\mu \rightarrow {x'}^\mu \right) 
    \theta^\gamma \theta^\delta = 
    \theta^\gamma \theta^\delta + 
    \varepsilon^\gamma \theta^\delta - 
    \varepsilon^\delta \theta^\gamma .
\label{sec3.2-12d}
\end{equation}
This equation has the following solution 
\begin{equation}
\begin{split}
    \delta\hat{B}\left(x^\mu \rightarrow {x'}^\mu \right) =  
    1 + &\varepsilon^\alpha \frac{\partial}{\partial \theta^\alpha} - 
    i\varepsilon^\alpha \sigma^\mu_{\alpha\dot{\beta}} 
    \bar{\theta}^{\dot{\beta}} \partial_\mu - \\
    &\bar{\varepsilon}^{\dot\alpha} 
    \frac{\partial}{\partial \bar{\theta}^{\dot\alpha}} + 
    i\theta^\alpha \sigma^\mu_{\alpha\dot{\beta}} 
    \bar{\varepsilon}^{\dot{\beta}} \partial_\mu
\label{sec3.2-12e}
\end{split}
\end{equation}
For the proof, see Appendix \ref{app}.
\par 
After this we can say that 
$\theta = \{ \theta^\alpha , \theta^{\dot\alpha} \}$ 
are the Grassmanian numbers which we should use as some additional 
coordinates for the description of the above-mentioned indefiniteness 
inherent at every spacetime point. 
\par 
Such approach can give us an excellent possibility for understanding 
of geometrical meaning of spin-$\hbar /2$. Wheeler \cite{wheel} 
has mentioned repeatedly the importance of a geometrical interpretation 
of spin-$\hbar /2$. He wrote: \textit{the geometrical description 
of $\hbar /2$-spin must be a significant component of any electron model} ! 
In this connection it should be note that Friedman and Sorkin 
\cite{friedman} for the first time have shown that the three manifold 
with non-trivial topology can have the quantum states of the gravitational 
field with half integral angular momentum. 

\section{Geometrical interpretation}

\begin{figure}
\begin{center}
\fbox{
\includegraphics[height=5cm,width=5cm]{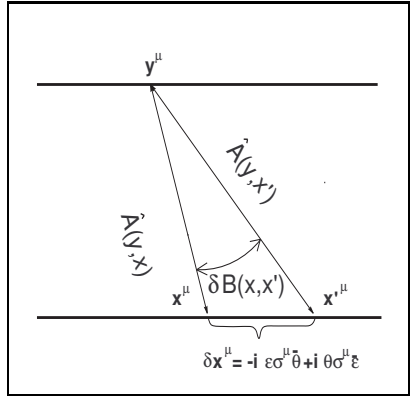}}
\caption{The distinction between two identification prescriptions
$\theta$ and $\theta'$ leads to a displacement $\delta x^\mu =
-i\varepsilon^\alpha \sigma^\mu_{\alpha\dot{\beta}}
  \bar{\theta}^{\dot{\beta}} +
  i\theta^\alpha \sigma^\mu_{\alpha\dot{\beta}}
  \bar{\varepsilon}^{\dot{\beta}}$}
\label{fig4}
\end{center}
\end{figure}

In this approach \textbf{superspace is an effective model of the spacetime
foam, \textit{i.e.} some approximation in quantum gravity}. The
indefiniteness connected with the creation/annihilation of quantum
minimalist wormholes is described by Grassmannian coordinates (see,
Fig.\ref{fig3}). In this interpretation an infinitesimal
Grassmannian coordinate transformation is associated with 
a displacement of the wormhole mouth, \textit{i.e.} with a 
change of the identification procedure (see, Fig.\ref{fig4})
\begin{equation}
\begin{split}
  &\theta' = \theta + \varepsilon =
  Id\biggl( y^\mu \;
  \raisebox{1.5ex}{$\underleftrightarrow{
  \scriptstyle{\;\;yes \; or \; no \; ?\;\;}}$}
  \;
  {x'}^\mu = \\
  &x^\mu - i\varepsilon^\alpha \sigma^\mu_{\alpha\dot{\beta}}
  \bar{\theta}^{\dot{\beta}} +
  i\theta^\alpha \sigma^\mu_{\alpha\dot{\beta}}
  \bar{\varepsilon}^{\dot{\beta}} 
  \biggl) = \hat{A}\left( y^\mu , {x'}^\mu \right) 
\label{sec1-20}
\end{split}
\end{equation}
here the identification procedure $Id(\ldots)$ 
is described by the operator 
$\hat{A}\left( y^\mu , {x'}^\mu \right)$. 
In this case the Grassmannian coordinate transformation
has a very clear geometrical sense : it describes a displacement 
of the wormhole mouth or it is the change of the
identification prescription. It is necessary to note that in 
Ref. \cite{gozzi} there is a similar interpretation of the 
Grassmanian ghosts : they are the Jacobi fields which are the 
infinitesimal displacement between two classical trajectories. 
\par
In this geometrical approach supersymmetry means that a
supersymmetrical Lagrangian is invariant under the identification
procedure \eqref{sec1-10}, \textit{i.e.}
\textbf{\textit{the corresponding supersymmetrical
fields must be described in an invariant manner on the background of
the spacetime foam.}}

\section{Acknowledgment}
I am very grateful for Doug Singleton and Richard Livine 
for the comments and fruitful discussion.

\appendix
\section{Calculation of $\delta\hat{B}$ operator}
\label{app}

For the proof of Eq. \eqref{sec3.2-12e} we shall calculate 
an effect of the $\delta B$ operator on the $\hat{A}$ operator. 
On the one hand we have 
\begin{equation}
\begin{split}
  &\delta\hat{B}\left(x^\mu \rightarrow {x'}^\mu \right) 
  \hat{A}^{\gamma\delta} \left( y^\mu , x^\mu \right) = \\
  &\biggl(1 + \varepsilon^\alpha \frac{\partial}{\partial \theta^\alpha} - 
    i\varepsilon^\alpha \sigma^\mu_{\alpha\dot{\beta}} 
    \bar{\theta}^{\dot{\beta}} \partial_\mu - \\
  &  \bar{\varepsilon}^{\dot\alpha} 
    \frac{\partial}{\partial \bar{\theta}^{\dot\alpha}} + 
    i\theta^\alpha \sigma^\mu_{\alpha\dot{\beta}} 
    \bar{\varepsilon}^{\dot{\beta}} \partial_\mu
  \biggl)
  \theta^\gamma \theta^\delta = \\
  &\theta^\gamma \theta^\delta + 
  \varepsilon^\gamma \theta^\delta - 
  \varepsilon^\delta \theta^\gamma \approx \\
  &\left( \theta^\gamma + \varepsilon^\gamma \right)
  \left( \theta^\delta + \varepsilon^\delta \right) = 
  {\theta '}^\gamma {\theta '}^\delta 
\label{app1}
\end{split}
\end{equation}
here ${\theta '}^\alpha = \theta^\alpha + \varepsilon^\alpha$. 
On the other hand 
\begin{equation}
\begin{split}
  &\delta\hat{B}\left(x^\mu \rightarrow {x'}^\mu \right) 
  \hat{A}^{\gamma\delta} \left( y^\mu , x^\mu \right) = \\
  &\varepsilon^{\gamma\delta} 
  \hat{A}\left( y^\mu , 
  \delta\hat{B}\left(x^\mu \rightarrow {x'}^\mu \right) x ^\mu \right) = \\
  &\varepsilon^{\gamma\delta} 
  \hat{A}\biggl( y^\mu ,
  \Bigl(
  1 + \varepsilon^\alpha \frac{\partial}{\partial \theta^\alpha} - 
    i\varepsilon^\alpha \sigma^\mu_{\alpha\dot{\beta}} 
    \bar{\theta}^{\dot{\beta}} \partial_\mu - \\
  &  \bar{\varepsilon}^{\dot\alpha} 
    \frac{\partial}{\partial \bar{\theta}^{\dot\alpha}} + 
    i\theta^\alpha \sigma^\mu_{\alpha\dot{\beta}} 
    \bar{\varepsilon}^{\dot{\beta}} \partial_\mu
  \Bigl)x^\mu
  \biggl) = \\
  &\varepsilon^{\gamma\delta} 
  \hat{A}\left( y^\mu , x^\mu - 
  i\varepsilon^\alpha \sigma^\mu_{\alpha\dot{\beta}} 
  \bar{\theta}^{\dot{\beta}} + 
  i\theta^\alpha \sigma^\mu_{\alpha\dot{\beta}} 
    \bar{\varepsilon}^{\dot{\beta}} = {x'}^\mu 
  \right) = \\ 
  &\hat{A}^{\gamma\delta} \left( y^\mu , x^\mu + 
  \delta x ^\mu\right)
\label{app2}
\end{split}
\end{equation}
where 
\begin{equation}
  \delta x ^\mu = 
  -i\varepsilon^\alpha \sigma^\mu_{\alpha\dot{\beta}} 
  \bar{\theta}^{\dot{\beta}} + 
  i\theta^\alpha \sigma^\mu_{\alpha\dot{\beta}} 
  \bar{\varepsilon}^{\dot{\beta}} . 
\label{app2a}
\end{equation}
Therefore we have 
\begin{eqnarray}
  &&\hat{A}^{\gamma\delta} 
  \left( y^\mu , {x'}^\mu = x^\mu + \delta x^\mu \right) = 
  \nonumber\\
  &&{\theta'}^\gamma {\theta'}^\delta = 
  \left( \theta^\gamma + \varepsilon^\gamma \right)
  \left( \theta^\delta + \varepsilon^\delta \right) \approx 
  \nonumber \\
  &&\theta^\gamma \theta^\delta + 
  \varepsilon^\gamma \theta^\delta - 
  \varepsilon^\delta \theta^\gamma . 
\label{app3}
\end{eqnarray}
It means that the operator 
$\hat{A}^{\gamma\delta}( y^\mu , {x'}^\mu )$ with the shifted 
wormhole mouth is equivalent to the change of the Grassmanian 
coordinate $\theta^\alpha \rightarrow {\theta'}^\alpha$.

\end{document}